# Волны плотности, додекаэдрическая геометрия и структуры некоторых сферических вирусных капсидов

О.В. Коневцова[1], С.Б. Рошаль[1] and В.Л. Лорман[2]

[1]*Физический факультет, Южный Федеральный Университет, ул. Зорге 5,г. Ростов-на-Дону, Россия, 344090.*

[2]*Лаборатория им. Шарля Кулона, CNRS – Университет Монпелье-2, пл. Е. Батайон, 34095, Монпелье, Франция*

Пентагональный порядок, образуемый 360 протеинами в капсиде вируса бычьей папилломы, и структуры некоторых других меньших вирусных капсидов рассматриваются с единой точки зрения, основанной на теории кристаллизации Л.Д. Ландау. Чтобы смоделировать устройство капсидов, мы отображаем на сферу через развертки додекаэдров разного размера плоскую пентагональную квазирешетку Пенроуза (ПКП), построенную в рамках теории Л.Д. Ландау. Рассмотрено шесть различных простейших додекаэдрических разверток, содержащих от 60 до 360 позиций, занимаемых протеинами. Появление кривизны у граней капсида и условие гладкого сопряжения распределения протеинов на смежных пятиугольных гранях, индуцируют двухкомпонентную фазонную деформацию квазикристаллического порядка на всех 12 гранях сферического додекаэдра. В результате данной самосогласованной перестройки образуются хиральные сферические структуры, расположение позиций в которых близко к расположению центров масс протеиновых молекул в известных вирусных капсидах. Предложенная теория объясняет общие закономерности организации протеинов в капсидах малых и средних вирусов.

## 1. Введение

Вирусы представляют собой микроскопические частицы, состоящие из белковой капсулы — капсида, и заключающегося в ней генома вируса. Закодированную в геноме генетическую информацию вирусы не могут самостоятельно реализовать, не обладая собственным механизмом синтеза белка. Вирусная оболочка, состоящая в большинстве случаев из множества идентичных копий одной асимметричной белковой молекулы, защищает и одновременно транспортирует геном к подходящей для заражения клетке. Инфицирование клетки-хозяина вирусом сильно зависит от расположения белков в капсиде [1-3]. Данное расположение с геометрической точки зрения было впервые интерпретировано Каспаром и Клугом (КК) в 1962 [4]. Согласно теории КК структуры значительного количества вирусных капсидов интерпретируются на основе развертки икосаэдра, треугольные грани которого заполняются периодическим гексагональным мотивом [4,5]. Особенности модели КК накладывают определенные ограничения на возможное общее количество белков в капсиде. В первую очередь данные ограничения связаны с икосаэдрической симметрией капсида. Асимметричные протеины могут расположиться только в 60-кратных позициях группы вращения икосаэдра *I*, поэтому общее число белковых молекул в капсиде равно 60T. И только величины $T=h^2+k^2+hk$, где h и k - неотрицательные целые числа, разрешены по правилам отбора КК.

Однако экспериментальные данные свидетельствуют о существовании вирусных капсидов, хотя и обладающих симметрией группы *I*, но неудовлетворяющих теории КК и соответствующих либо запрещенным числам *T*, либо характеризуемые локальным порядком протеинов, не представимым в виде гексамеров и пентамеров. В недавней работе [6] был рассмотрен один из таких капсидов, а именно капсид вируса бычьей папилломы (БПВ), и была предложена его структурная модель впервые основанная на додекаэдрической развертки капсида. Только такой тип развертки в принципе был совместим с пентагональным порядком протеинов, наблюдаемом в капсиде БПВ. В предложенной модели [6] каждая из граней развертки капсида была декорирована хиральным образом перестроенной пентагональной квазирешеткой Пенроуза (ПКП) [7]. Протеины располагались в вершинах тайлинга, состоящего из трех типов тайлов - пятиугольников, узких и широких ромбов. Минимальный набор тайлов, представленных в структуре, делал ее совместной с принципом квазиэквивалентности, а ее практической реализации в природе способствовал тот факт, что организовать укладку из 360 протеинов в рамках геометрической модели Каспара-Клуга просто невозможно. Важной особенностью работы [6] был ее физический, а не чисто геометрический характер. В продолжение хорошо известных работ [8,9] в [6] вирусная структура рассматривалась с обычных для физики конденсированного состояния позиций, и хиральность ПКП, декорирующей грани развертки капсида выводилась путем минимизации фонон-фазонной упругой энергии квазикристаллического порядка. Фактически хиральность индуцировалась за счет разрешенного по симметрии билинейного взаимодействия между обычной (фононной) деформацией выпучивания капсида и фазонной деформацией.

В более ранних работах [8,9] структуры некоторых других малых сферических вирусов, как удовлетворяющие, так и не удовлетворяющие геометрической конструкции КК были выведены в рамках подхода волн плотности и минимизация функционала свободной энергии Ландау. Основной целью данной работы будет обобщение и объединение предложенных нами ранее для разных вирусных капсидов казалось бы, совершенно разных физических и геометрических концепций, таких как подход волн плотности и теория сферической кристаллизации Ландау [8] с одной стороны и додекаэдрическая развертка структуры капсида и минимизация фонон-фазонной упругой энергии его граней, декорированных квазикристаллическим порядком [6], с другой стороны. При этом в данной работе будет продемонстрирована универсальность теории кристаллизации Ландау и вытекающего из нее подхода волн плотности. Мы выведем структуру БПВ непосредственно в рамках подхода волн плотности. При этом окажется, что данная структура описывается в рамках исключительно плоских волн плотности, рассматриваемых на поверхности додекаэдра. В то же время структуры меньших по числу протеинов капсидов, выведенные нами ранее в рамках подхода сферических волн плотности [8], могут быть также получены в рамках формализма плоских волн плотности, также рассматриваемых на поверхности додекаэдра.

Данная работа устроена следующим образом. В [6] было показано, что локальная организация протеинов в БПВ капсиде родственна устройству ПКП. Поэтому в следующем разделе мы кратко рассмотрим теорию кристаллизации и подход волн плотности для данной решетки. Также в рамках теории кристаллизации Ландау мы особо остановимся на возможном в данной решетке хиральном билинейном фонон-фазонном взаимодействии. Такое взаимодействие способно индуцировать хиральный квазикристаллический порядок за счет неоднородной, но ахиральной обычной деформации. Третий раздел посвящен выводу в рамках теории плоской кристаллизации Ландау и подхода волн плотности структуры БПВ и некоторых меньших сферических капсидов. В этом же разделе будет предложена новая структурная модель сдвоенного капсида вируса кукурузного стрика из семейства Близнецов [10]. Выводы и заключение будут изложены в последнем четвертом разделе.

## 2. Образование и фазонная деформация пентагональной квазирешетки Пенроуза с точки зрения теории кристаллизации Ландау

Рассмотрим некоторые основные особенности теории кристаллизации Ландау [11] применительно к плоским декагональным квазикристаллам [12]. Вблизи точки кристаллизации распределение плотности структурных единиц может быть записано в следующем виде

$$\rho(\mathbf{R}) = \rho_0 + \delta\rho(\mathbf{R}), \qquad (1)$$

где $\rho_0$ - кластерная плотность до кристаллизации, а $\delta\rho(\mathbf{R})$ соответствует отклонению критической плотности, вызванному образованием квазикристаллического порядка. Согласно теории Ландау $\delta\rho(\mathbf{R})$ - неприводимая функция, включающая единичное неприводимое представление группы симметрии изотропного состояния. Соответствующее разложение функции $\delta\rho(\mathbf{R})$ по плоским волнам принимает вид:

$$\delta\rho(\mathbf{R}) = \sum_{k=0}^{9} \rho_k \exp(i\mathbf{b}_k \mathbf{R}), \qquad (2)$$

где $\mathbf{R}$ – радиус вектор, $\mathbf{b}_k = b^0 \langle \cos(k\pi/5), \sin(k\pi/5) \rangle$. Так как $\rho_k = |\rho_k| \exp(i\phi_k^0)$ и отклонение плотности $\delta\rho(\mathbf{R})$ действительны, $\rho_\mathbf{k} = \rho_j^*$ для $j=(k+5)$ mod 10 и число неприводимых (независимых) фаз $\phi_k^0$ равно 5. Разложение свободной энергии Ландау может быть представлено как инвариантная функция амплитуд $\rho_\mathbf{k}$. Эта энергия зависит лишь только от симметричной комбинации $\xi = \sum_{n=0}^{4} \phi_{2n}^0$ и не зависит от 4 других ортогональных линейных комбинаций фаз $\phi_k^0$ [13]. Минимизация свободной энергии Ландау относительно $|\rho_\mathbf{k}|$ и $\phi_k^0$ позволяет определить лишь сумму фаз, но не дает значений фаз непосредственно. Так происходит потому, что в данном случае квазикристаллическая функция плотности $\rho(\mathbf{r})$ раскладывается в ряд Фурье с минимальным числом $N$ базисных векторов большим, чем размерность $n$ квазикристалла в физическом прямом пространстве (в рассматриваемом случае $N=4$ и $n=2$). Однородные сдвиги в пятимерном фазовом пространстве $\{\phi_k^0\}$, изменяющие сумму фаз, изменяют в свою очередь и энергию. Оставшиеся четыре степени свободы в фазах могут быть параметризованы через два двумерных вектора: вектор смещений, соответствующий сдвигу квазикристалла как целого - $\mathbf{u}$ (этот же вектор появляется и в теории кристаллизации периодической решетки) и фазонный вектор - $\mathbf{v}$, отсутствующий в случае периодическом случае и являющийся ортогональным дополнением к физическому пространству. Векторы $\mathbf{u}$ и $\mathbf{v}$ преобразуются по двум двумерным векторным неприводимым представлениям точечной группы симметрии декагонального квазикристалла. Математически переменная $\mathbf{v}$ аналогична фазонной степени свободы в несоразмерных кристаллах. В случае декагональной симметрии параметризация фаз имеет вид [13]:

$$\phi_{2n}^0 = \mathbf{b}_{2n}\mathbf{u} + \mathbf{b}_{2n}^\perp \mathbf{v} + \xi/5, \qquad (3)$$

где $\mathbf{b}_{2n}^\perp = b^0 \langle \cos(6n\pi/5), \sin(6n\pi/5) \rangle$ - векторы перпендикулярного обратного пространства.

Симметрия десятого порядка приводит к равенству всех значений $|\rho_k|$ и функция $\Delta\rho(\mathbf{R})$ переписывается в следующем действительном виде

$$\Delta\rho(\mathbf{r}) = 2\rho_\Delta \sum_{n=0}^{4} \cos(\mathbf{b}_{2n}\mathbf{r} + \phi_{2n}^0), \qquad (4)$$

где, $\rho_\Delta = |\rho_k|$.

Структурные единицы в получаемой квазирешетке размещаются в наиболее интенсивных *максимумах* функции плотности (4). Именно такой способ размещения структурных единиц минимизирует энергию системы [14]. С очень хорошей точность [14] координаты максимумов определяются как

$$\mathbf{r}_j = \sum_{i=0}^{4} N_{2i}^j \mathbf{a}_i - \mathbf{u}, \qquad (5)$$

где $\mathbf{a}_i = \dfrac{4\pi}{5b_0}(\cos(i2\pi/5),\sin(i2\pi/5))$, $i=0,1,2,3,4$. $N_{2i}^j$ - целые числа. Ортогональными дополнение к (5) будут подпространства

$$\mathbf{r}_j^\perp = \sum_{i=0}^{4} N_{2i}^j \mathbf{a}_i^\perp - \mathbf{v}, \qquad (6)$$

$$\xi = 2\pi \sum_{i=0}^{4} N_{2i}^j. \qquad (7)$$

где $\mathbf{a}_i^\perp = \dfrac{4\pi}{5b_0}(\cos(i6\pi/5),\sin(i6\pi/5))$, $i=0,1,2,3,4$. Наибольшая интенсивность максимума примерно определяется [14] длиной вектора (6) как

$$\delta\rho(\mathbf{r}_j) \approx 10\rho_\Delta \left(1 - (\mathbf{r}_j^\perp)^2 b_0^2/4\right). \qquad (8)$$

Иными словами, чем больше интенсивность максимума, тем меньше длина вектора (6). Энергия квазикристаллической решетки зависит от (7). Энергетическому минимуму соответствует случай, когда величина $\xi_0 = \xi/(2\pi)$ является целой или полуцелой [14].

Все максимумы функции плотности (4) в области $\delta\rho(\mathbf{r}_j) > 3.08\rho_\Delta$ индексируется целочисленными индексами $N_{2n}^j$ [14]. В то же время функция плотности (4) обрезанная любым подобным образом, содержит дополнительные максимумы, не принадлежащие идеальной пентагональной укладке Пенроуза, для построения которой нужны дополнительные правила отбора узлов. Их можно либо сформулировать в рамках условной минимизации свободной энергии [14], либо в рамках эквивалентной концепции проекционного окна [15], согласно которой узел принадлежит решетке, если его перпендикулярные координаты попадают внутрь проекционного окна. Однако, для применения теории кристаллизации к вирусным капсидам, дополнительные правила отбора узлов не очень существенны из-за ограниченного размера грани капсида. Достаточно лишь использовать подходящее по высоте обрезание функции $\delta\rho(\mathbf{R})$, эквивалентное проекционному окну в форме круга. Также заметим, что если присутствующий в выражении (6) вектор $\mathbf{v}$, соответствующий однородному фазонному сдвигу, является нулевым, то укладка обладает глобальной осью пятого порядка. Укладки с величинами $\xi_0$ различающимися на 5 совпадают. Если $\xi_0 \neq 0$ локальные оси не проходит через узлы решетки, что делает подобные укладки пригодными для декорирования асимметричными протеинами. При этом протеины в развиваемом подходе могут занимать непосредственно узлы квазирешетки, а не позиции общего положения в окрестности узлов, как в случае капсидов с локальным гексагональным порядком [4].

Рассмотрим теперь неоднородную фазонную деформации ПКП в рамках которой удалось объяснить хиральный порядок протеинов в капсиде БПВ [6]. Как было отмечено ранее, если величины $\mathbf{v}$ и $\mathbf{u}$ изменяются однородно во всем объеме, то свободная энергия решетки остается инвариантной. При этом решетка испытывает либо переключения узлов, соответствующие однородному фазонному сдвигу, либо смещается в пространстве как целое. При неоднородном изменении данных величин ПКП искажается, а ее свободная энергия получает соответствующую добавку. Неоднородное изменение величины $\mathbf{u}$ соответствует обычной упругой деформации. Неоднородное изменение величины $\mathbf{v}$ называется фазонной деформацией. Плотность фонон-фазонной упругой энергии квазирешетки с хиральной симметрией $C_5$, представляющая собой добавку к свободной энергии Ландау (зависящей от амплитуд волн плотности) выражается как инвариантная квадратичная функция пространственных производных двух полей $\mathbf{u}$ и $\mathbf{v}$. Построить эту функцию удобнее всего в формально-комплексном виде. Для этого заметим, что под действием поворота на $2\pi/5$ пара комплексно-сопряженных функций $U_1 = u_x + iu_y$ и $U_2 = u_x - iu_y$ просто умножаются на коэффициенты $\exp(-2\pi i/5)$ и $\exp(2\pi i/5)$, соответственно. Аналогично ведут себя дифференциальные операторы $\Delta_1 = \partial_x + i\partial_y$ и $\Delta_2 = \partial_x - i\partial_y$. Поле **v** преобразуется по другому неприводимому представлению, и под действием поворота на $2\pi/5$ пара комплексно-сопряженных функций $V_1 = v_x + iv_y$ и $V_2 = v_x - iv_y$ умножается на коэффициенты $\exp(-6\pi i/5)$ и $\exp(6\pi i/5)$. Поэтому нетривиальные (отсутствующие в обычной двумерной сплошной среде) фазонные и фонон-фазонные квадратичные инварианты можно записать в следующем виде: $(\Delta_1 V_1)(\Delta_2 V_2)$; $(\Delta_1 V_2)(\Delta_2 V_1)$; $(\Delta_1 U_1)(\Delta_1 V_2) + (\Delta_2 U_2)(\Delta_2 V_1)$; $(\Delta_1 U_1)(\Delta_1 V_2) - (\Delta_2 U_2)(\Delta_2 V_1)$. Окончательно, после раскрытия скобок и некоторых переобозначений, плотность фонон-фазонной упругой энергии может быть записана в следующем виде

$$F = \frac{1}{2}\lambda(\varepsilon_{ii})^2 + \mu\varepsilon_{ij}\varepsilon_{ij} + \frac{1}{2}K_1(\partial_i v_j)(\partial_i v_j) + K_2\big((\partial_x v_x)(\partial_y v_y) - (\partial_y v_x)(\partial_x v_y)\big) + \\ + K_3\big[(\varepsilon_{xx} - \varepsilon_{yy})(\partial_x v_x + \partial_y v_y) + 2\varepsilon_{xy}(\partial_x v_y - \partial_y v_x)\big] + \\ + K_4\big[2\varepsilon_{xy}(\partial_x v_x + \partial_y v_y) - (\varepsilon_{xx} - \varepsilon_{yy})(\partial_x v_y - \partial_y v_x)\big] \quad , \quad (9)$$

где $i=x,y$; $\varepsilon_{ij} = (\partial_j u_i + \partial_i u_j)/2$. Хиральной симметрии $C_{10}$ соответствует упругая энергия той же функциональной формы.

Впервые фонон-фазонная упругая энергия в форме (9) была получена в [13] как энергия, соответствующая нехиральной группе симметрии $C_{10v}$. Однако, последующий анализ [13] показал, член при $K_4$ не инвариантен относительно действия всей группы симметрии $C_{10v}$, но становится таковым только относительно ее хиральной подгруппы с симметрией $C_5$ или $C_{10}$. Таким образом, член при $K_4$ фактически является псевдоскаляром. Подобная ситуация реализуется, например, в теории хиральных жидких кристаллов [16,17], где дополнительный псевдоскалярный член (**n curl n**) линейный по первым производным по направлению **n** появляется в упругой свободной энергии Франка-Озеена холестерика. Однако, любая даже *хиральная квазикристаллическая симметрия запрещает существование чисто фазонных псевдоскалярных членов, состоящих исключительно из первых пространственных производных поля* **v**. Видимо этот факт объясняет исключительность хирального квазикристаллического порядка в природе, обнаруженного пока исключительно в вирусных капсидах. Индуцировать линейным образом (без фазового перехода) в рамках хиральной квадратичной упругой энергии (типа (9)) хиральный квазикристаллический порядок может только неоднородная деформация. Подобная линейная индукция в рассматриваемом случае возможна исключительно через член при коэффициенте $K_4$. Именно этот член индуцирует хиральность структуры вирусных капсидов, рассматриваемых в следующем разделе. Изложение материала данного раздела мы начнем со структуры капсида БПВ, исследованной нами ранее в [6]. Однако в отличие от [6] структуры рассматриваемых капсидов будут описываться с привлечением формализма волн плотности.

### 3. Описание структура БПВ и некоторых других капсидов в рамках формализма волн плотности

Следуя [6], мы предполагаем, что хиральность расположения узлов квазирешетки индуцируется нелинейной деформацией граней капсида, возникающей при их выпучивании. Тогда, для простоты полагая окончательную форму капсида сферической, а поле смещений – радиальным, можно вычислить тензор деформации соответствующего выпучивания. Затем, минимизируя упругую энергию квазикристалла относительно фазонных переменных, можно получить выражение для фазонного поля, приводящего к хиральности.

Предположение о радиальном выпучивании позволяет связать координаты радиус вектора **R**, лежащего на сферическом сегменте радиуса R, с координатами его проекции

**R**'=<$x,y,h$> на плоскую грань, находящуюся на расстоянии $h$ от центра сферы. Выбирая для определенности, грань капсида, перпендикулярную направлению Z получаем:

$$\mathbf{R} = \frac{R}{\sqrt{x^2 + y^2 + h^2}} <x, y, h>. \tag{10}$$

Соответствующий тензор деформации тогда $\boldsymbol{\varepsilon} = \frac{1}{2}(\mathbf{M}_s - \mathbf{M}_p)$ [17], где

$$\mathbf{M}_s = \frac{R^2}{(x^2 + y^2 + h^2)^2} \begin{bmatrix} y^2 + h^2 & xy \\ xy & x^2 + h^2 \end{bmatrix} \tag{11}$$

метрический тензор сферического сегмента, а $\mathbf{M}_p$ - метрический тензор, соответствующий плоскости (т.е. единичная матрица). Явная форма члена в плотности упругой энергии, выражающего связь фононной и фазонной подсистемы хирального квазикристалла, получается в результате подстановки тензора деформации $\varepsilon$ в член при коэффициенте $K_4$ выражения (9). Предполагая, что кривизна является слабой (или h≈R) и, разлагая тензор $\varepsilon$ в ряд с точностью до первых неисчезающих членов по $x$ и $y$, получаем инвариант при $K_4$ (смотрите (9)) в следующем виде:

$$J_4 = \frac{2xy(\partial_x v_x + \partial_y v_y) + (x^2 - y^2)(\partial_y v_x - \partial_x v_y)}{2R^2} \tag{12}$$

В данном рассмотрении мы ограничимся минимальной упругой фонон-фазонной энергией, приводящей к хиральному квазикристаллическому порядку. Два чисто фононных члена энергии (9) учитывать бессмысленно, так как тензор деформации в рассматриваемом приближении считается заранее заданным. Нехиральный фонон-фазонный член исключается для упрощения задачи, так как он индуцирует нехиральное фазонное поле и к хиральному квазикристаллическому порядку привести не может. Окончательно в исследуемой энергии помимо члена, пропорционального (12), остается еще два слагаемых:

$$F = \int_S \left( \frac{1}{2} K_1 J_1(\mathbf{v}) + K_2 J_2(\mathbf{v}) + K_4 J_4(\mathbf{v}) \right) dS, \tag{13}$$

где $J_1 = (\partial_x v_x)^2 + (\partial_y v_x)^2 + (\partial_x v_y)^2 + (\partial_y v_y)^2$ и $J_2 = (\partial_x v_x)(\partial_y v_y) - (\partial_y v_x)(\partial_x v_y)$. Функционал энергии (13) минимизируется в пределах круга с радиусом $r_0$, где $r_0$ - эффективный радиус грани капсида. При этом получается как вид искомого поля:

$$v_1^0 = \beta (y^3/3 - x^2 y); \quad v_2^0 = \beta (x^3/3 - xy^2), \tag{14}$$

так и значение параметра $\beta$, представляющего собой коэффициент хиральности:

$$\beta = \frac{K_3}{2R^2(K_1 + K_2)}. \tag{15}$$

Соответствующая минимуму фонон-фазонная энергия записывается в следующем виде:

$$F_0 = -\frac{\pi r_0^6 K_3^2}{12 R^4 (K_1 + K_2)}. \tag{16}$$

Заметим, что поле (14) в принципе определено с точностью до постоянного члена, не зависящего от пространственных координат. Однако, чтобы глобальная ось симметрии деформированной структуры проходила через начало координат, данный член следует положить равным нулю.

Фазонные поля, приводящие к хиральной квазикристаллической симметрии и сохраняющие поворотную ось старшего порядка, можно также получить с помощью теории групп. Для этого вместо компонент радиус вектора $x$ и $y$ удобно ввести их линейные комплексно-сопряженные комбинации $r_1 = x - Iy$ и $r_2 = x + Iy$, преобразующиеся по тому же представлению, что и введенные выше дифференциальные операторы $\Delta_1 = \partial_1 + i\partial_2$ и $\Delta_2 = \partial_i - i\partial_2$. Чтобы получить искомые зависимости $\mathbf{v}(\mathbf{r})$, приводящие к образованию хиральных структур, достаточно построить формально комплексные псевдоскаляры из величин $\mathbf{v}$ и $\mathbf{r}$, линейные по $\mathbf{v}$. Два таких первых псевдоскаляра приведены ниже:

$$S_1 = (v_1 - Iv_2)(x - Ix)^2 - (v_1 + Iv_2)(x + Iy)^2 \qquad (17)$$

$$S_2 = (v_1 - Iv_2)(x + Iy)^3 - (v_1 + Iv_2)(x - Iy)^3 \qquad (18)$$

Форма поля (14) получается простым дифференцированием псевдоскаляра $S_2$ по компонентам $\mathbf{v}$. Дифференцирование остальных псевдоскаляров даст фазонные поля, сохраняющие осевую симметрию, но, естественно, не минимизирующие энергию (13).

Однако, как было продемонстрировано в [6] структура капсида БВП не может быть выведена за счет включения исключительно поля (14) и энергия квазирешетки должна минимизироваться с дополнительным граничным условием. Данное граничное условие благоприятствует склейке смежных граней и состоит во включении в поле фазонных деформаций (14) простейшей полносимметричной компоненты. Найти явный вид такой компоненты несложно, записав простейший по форме скаляр величин $\mathbf{v}$ и $\mathbf{r}$ линейный по $\mathbf{v}$:

$$S_1 = (v_1 - Iv_2)(x - Ix)^2 + (v_1 + Iv_2)(x + Iy)^2 \qquad (19)$$

Дифференцирование (19) по $v_1$ и $v_2$ задает форму искомого поля

$$v_1^1 = \alpha(y^2 - x^2); \quad v_2^1 = 2\alpha xy, \qquad (20)$$

где коэффициент $\alpha$ определяет амплитуду данного ахирального поля. Заметим, что поле (20) включается исключительно склейкой смежных граней и не может быть получено за счет учета ахирального фонон-фазонного инварианта, исключенного из энергии (13).

Связь между ПКП и экспериментально наблюдаемым распределением белковых молекул в капсиде БПВ показана на рисунке 1 и может быть представлена как двухшаговое переключение нескольких позиций [6]. Каждый шаг затрагивает только одну орбиту группы симметрии $C_{5v}$ плоской квазирешетки. Такого рода переключения хорошо известны в классических квазикристаллах и обычно называются фазонными прыжками.

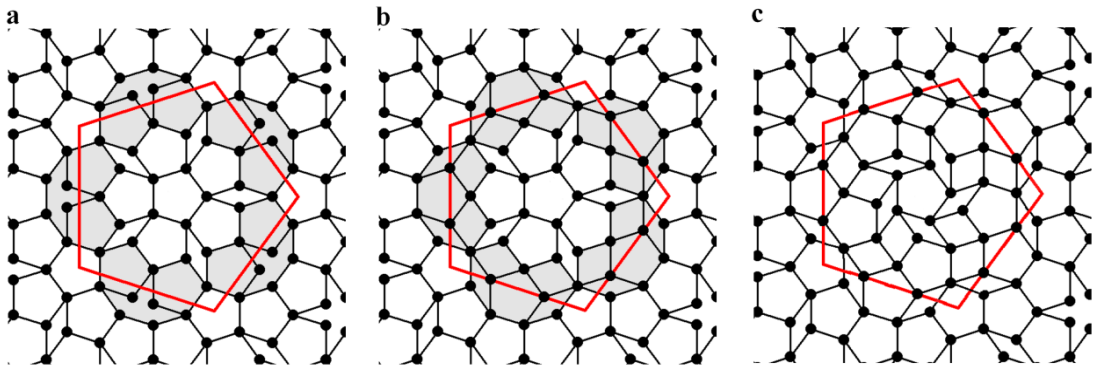

**Рисунок 1.** Пентагональный квазикристаллический порядок соответствующий одной пятиугольной грани сферического додекаэдра капсида вируса бычьей папилломы. Протеины представлены черными кругами. Грань додекаэдрической развертки показана большим красным пятиугольником. (a) - Обычная пентагональная квазирешетка Пенроуза с глобальной осью в центре панели. (b) - Квазирешетка после действия поля фазонных деформаций (20), ответственного за согласование пентагонального порядка на смежных гранях додекаэдра. Переключение позиций происходит в зоне, выделенной серым цветом. (c) - Окончательная форма квазирешетки после действия суммарного поля фазонных деформаций (21).

Поле (20) можно включить в суммарное фазонное поле $\mathbf{v}$ не только путем симметрийных рассуждений, но и путем прямой минимизации энергии (13) при определенных граничных условиях, а именно интеграл по замкнутому круглому контуру ($y^2+x^2=r_0^2$) от скалярного произведения ($\mathbf{vv}^1$) должен быть равен интегралу по тому же контуру от величины ($\mathbf{v}^1)^2$, где $\mathbf{v}^1$ - поле (20). Даже в этом сложном случае вариационная задача имеет точное решение:

$$v_1=\alpha(y^2-x^2)+\beta(y^3/3-x^2y); \quad v_2=2\alpha xy+\beta(x^3/3-xy^2) \tag{21}$$

Выражение (21) представляет собой общее поле фазонных деформаций для одной грани вирусного «наноквазикристалла». Величина параметра $\beta$ при включении граничных условий не меняется, а величина свободной энергии, соответствующей условному минимуму $F_0$, становится больше. Окончательная фонон-фазоннная энергия $F_0^1$ тогда может быть записана в следующем виде:

$$F_0^1 = -\frac{\pi\ r_0^6 K_3^2}{12R^4(K_1+K_2)} + 2(K_1-K_2)\alpha^2\pi\ r_0^4 \tag{22}$$

В работе [6] выражения (5-6) использовались совместно с полем (21) для расчета позиций фазонно-деформированной квазирешетки. При этом форма окна проектирования считалась неизменной в процессе фазонной деформации, а величина коэффициента хиральности $\beta$ связывалась с параметрами фонон-фазонной энергии (13) по формуле (15). Коэффициент $\alpha$ по-существу просто подбирался, чтобы обеспечить желаемый порядок протеинов в пределах грани. Оказалось, что структура капсида БПВ реализуется в широком диапазоне значений $\alpha$ и $\beta$, однако как-то более точно определить параметры поля (21) не удалось.

На самом деле фонон-фазонная упругая энергия - лишь часть полной свободной энергии Ландау, которая наряду с производными полей смещений по координатам также зависит от амплитуд волн плотности $\rho_k$ (2) в квазикристаллической фазе. Выделенность величин параметров $\alpha$ и $\beta$ должна обеспечиваться именно минимизацией полной свободной энергии. Реализовать подобную минимизацию практически – необозримая задача. Однако если считать, что часть сводной энергии Ландау, связанная с образованием квазирешетки (или с амплитудами волн плотности $\rho_k$), достаточно велика по сравнению с энергией фазонной деформацией, то определить приближенно область, где должны лежать параметры $\alpha$ и $\beta$ можно следующим довольно простым путем.

В работе [14] было предложено рассматривать квазикристаллическую фазу как предельную, то есть такую, минимальность свободной энергии Ландау в которой обеспечивается экстремальностью (насыщением) эффективного параметра порядка в фазе. Ниже мы используем туже самую идеологию для примерного расчета параметров $\alpha$ и $\beta$. Явный вид эффективного параметра порядка, ответственного за образование квазирешетки (без учета ее фазонной деформации) выводится из тех соображений, что параметр порядка должен быть полностью симметричен в низко симметричной фазе [4, 14]. Поэтому параметр порядка перехода изотропное состояние – декагональная фаза (без фазонной деформации) выражается как средняя амплитуда по амплитудам $\rho_k$:

$$\rho_\Delta = \frac{1}{5S}\sum_{n=0}^{4}\sum_{i=1}^{N}\cos(\mathbf{b}_{2n}\mathbf{r}_i + \phi_{2n}^0), \tag{23}$$

где фазы $\phi_{2n}^0$ параметризуются по формуле (3), S – площадь грани, N=35 – число частиц на грани, включая ее границу, $\mathbf{r}_i$ - координаты частиц. Подставим теперь в (3) $\mathbf{u}$=0 и запишем $\mathbf{v}$ в форме (21). При этом вместо радиус-вектора ($x,y$) следует брать координаты конкретных частиц $\mathbf{r}_i$. Максимизация полученного выражения по отношению к $\mathbf{r}_i$ и к величинам параметров $\alpha$ и $\beta$ соответствует минимизации части свободной энергии, которая не содержит пространственные производные. Одновременно этот процесс определяет как координаты частиц, так и величины параметров $\alpha$ и $\beta$. Для бесконечной квазирешетки минимум (23) соответствует нулевым значениям параметров $\alpha$ и $\beta$. Результирующее расположение узлов показано на рисунке 1(а). Для конечных областей квазирешетки у выражения (23) существует минимумы как типа ($\alpha=0$ и $\beta\neq 0$), один из которых соответствует ахиральному расположению частиц, принадлежащих большому пятиугольнику, показанному на рисунке 1(b), так и минимумы типа ($\alpha\neq 0$ и $\beta\neq 0$), один из которых соответствует хиральной структуре, лежащей в пределах пятиугольника, показанного на рисунке 1(c).

Так как протеины в рассматриваемой теории всегда находятся в максимумах (4), то для заранее заданного расположения частиц по максимумам (4) расчет параметров $\alpha$ и $\beta$ можно упростить, воспользовавшись выражением (8), где величина $\mathbf{r}_i^\perp$ определяется выражением (6), которое в свою очередь параметризуется (21). В таком случае максимизация (23) эквивалентна минимизации по $\alpha$ и $\beta$ выражения

$$\sum_{i=1}^{N}(\mathbf{r}_i^\perp(\mathbf{v}_i(\alpha,\beta)))^2, \qquad (24)$$

где $\mathbf{v}_i$ определяется полем (21), куда вместо координат радиус-вектора ($x,y$) следует подставить параллельные координаты конкретных частиц $\mathbf{r}_i$.

Для $N=35$ (где $N$ - число протеинов на грани БПВ капсида, включая ее границы) минимизация (24) дает $\alpha_0=0{,}0033$ и $\beta_0=-0{,}00005$. Данная точка ($\alpha_0,\beta_0$) попадает внутрь (но вблизи границы) области параметров $\alpha$ и $\beta$, при которых фазонная деформация приводит к структуре грани БПВ капсида. Для построения позиций этой структуры используются формулы проектирования (5-6), форма фазонной деформации (21), условие $\xi_0=1$, и проекционное окна радиусом $|\mathbf{r}^\perp|\approx 0.88$. Также для упрощения положено, что $b_0=1$ и $\mathbf{u}=0$. Тот факт, что структура БПВ капсида может соответствовать минимуму амплитудо-зависимой части потенциала Ландау говорит, что, обе составляющие потенциала (зависящая от амплитуд и зависящая от пространственных производных фаз) минимизируются при близких значениях $\alpha$ и $\beta$. Таким образом полный потенциал Ландау также обязан иметь минимум внутри данной области.

На рисунке (2) показана функция плотности (23) в области $\Delta\rho(\mathbf{r})>4\rho_\Delta$ после действия поля фазонных деформаций (21), где $\alpha_0=0.0033$ и $\beta_0=-0.00012$ (панель а) и структуры граней БПВ и некоторых других (меньших) капсидов (панель б). Выбранные для построения рисунка параметры $\alpha$ и $\beta$ примерно соответствуют середине области изменения этих параметров, соответствующей БПВ капсиду.

Интересно отметить, что в рамках изложенной выше теории можно получить и структуры других меньших капсидов, рассмотренные нами ранее в [8,9]. Пятиугольные грани соответствующих додекаэдрических разверток вырезаются из той же самой хиральной квазикристаллической структуры (см. рис. 2) способами, показанными на правой панели. Таким образом, хиральность порядка протеинов в данных капсидах также может объясняться выпучиванием их граней. То, что для разных капсидов, данная хиральная структура оказывается единой, означает, что все рассматриваемые капсиды, несмотря на разное количество белковых молекул, стремятся организовать похожие хиральные мотивы вокруг оси пятого порядка. Построенный тайлинг довольно закономерен и по чисто геометрическим причинам. Так, сопрячь на плоскости пять пентамеров с центральным можно только при помощи узких ромбов. Если для этой цели использовать широкие ромбы, то тайлинг из тех же самых тайлов (имеющихся исходно) просто не соберется.

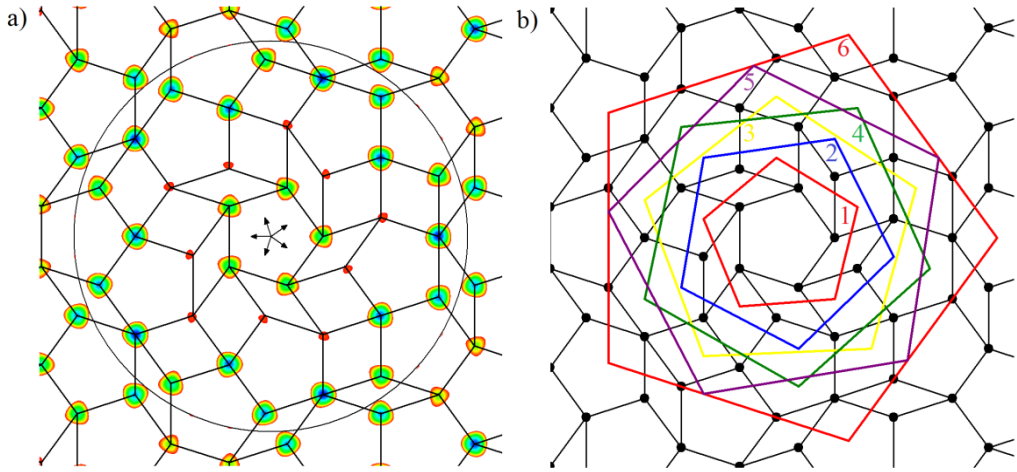

**Рисунок 2** - Функция плотности (23) в области $\Delta\rho(\mathbf{r}) > 4\rho_\Delta$ после действия поля фазонных деформаций (21) (панель a) и структуры граней БПВ и некоторых других капсидов (панель b). Изменение цветов от красного к фиолетовому (панель a) соответствует изменению величины функции $\delta\rho(\mathbf{R})$ в пределах от $4\rho_\Delta$ до $10\rho_\Delta$. Положения максимумов и положения проекций соответствующих узлов пятимерного пространства практически совпадают. Пять векторов в центре рисунка a) соответствуют двумерным базисным трансляциям $\mathbf{a}_i$ (5). Узлы, ближайшие к центру индексируются как (1,-1,1,0,0) + циклическая перестановка. Грани капсида БПВ соответствуют самому большому (внешнему) пятиугольнику, из показанных на рисунке 2(a). Пятиугольные грани додекаэдрических разверток других капсидов, обсуждаемых ниже, показаны большими пятиугольниками разных цветов, обозначенными цифрами от 1 до 6.

На рисунке 3 представлены развертки шести капсидов, соответствующие пятиугольным граням (1-6), представленным на рис. 2(b). Так же как и в случае капсида БПВ, рассмотренном в [6], для того, чтобы соответствовать глобальной икосаэдрической симметрии, двумерные додекаэдрические развертки капсидов (a-d) должны обладать осями симметрии второго порядка, проходящими через середины сторон пятиугольных граней. Ось такого типа в пятимерном пространстве E является суперпозицией инверсии и трансляции **P**, соединяющей центры соседних граней развертки. Данная трансляция **P** также представима в виде суммы пятимерных координат (**A** и **A′**) двух смежных вершин одной пятиугольной грани развертки. В таблице 1 приведены по одной трансляции типа **P** и **A** для каждой из показанных на рис. 3 разверток.

Таблица 1. Геометрические характеристики представленных на рисунке 3 разверток

|   | a) | b) | c) | d) | e) | f) |
|---|---|---|---|---|---|---|
| **P** | 2,3,0,-2,-1 | 2,4,1,-3,-2 | 5,3,-2,-4,0 | 3,5,1,-4,-3 | 6,4,-2,-5,-1 | 2,8,2,-5,-5 |
| **A** | 2,1,-1,-1,0 | 2,2,-1,-2,0 | 3,0,-2,-2,2 | 3,2,-1,-3,0 | 4,0,-2,-3,2 | 4,4,-2,-3,-2 |

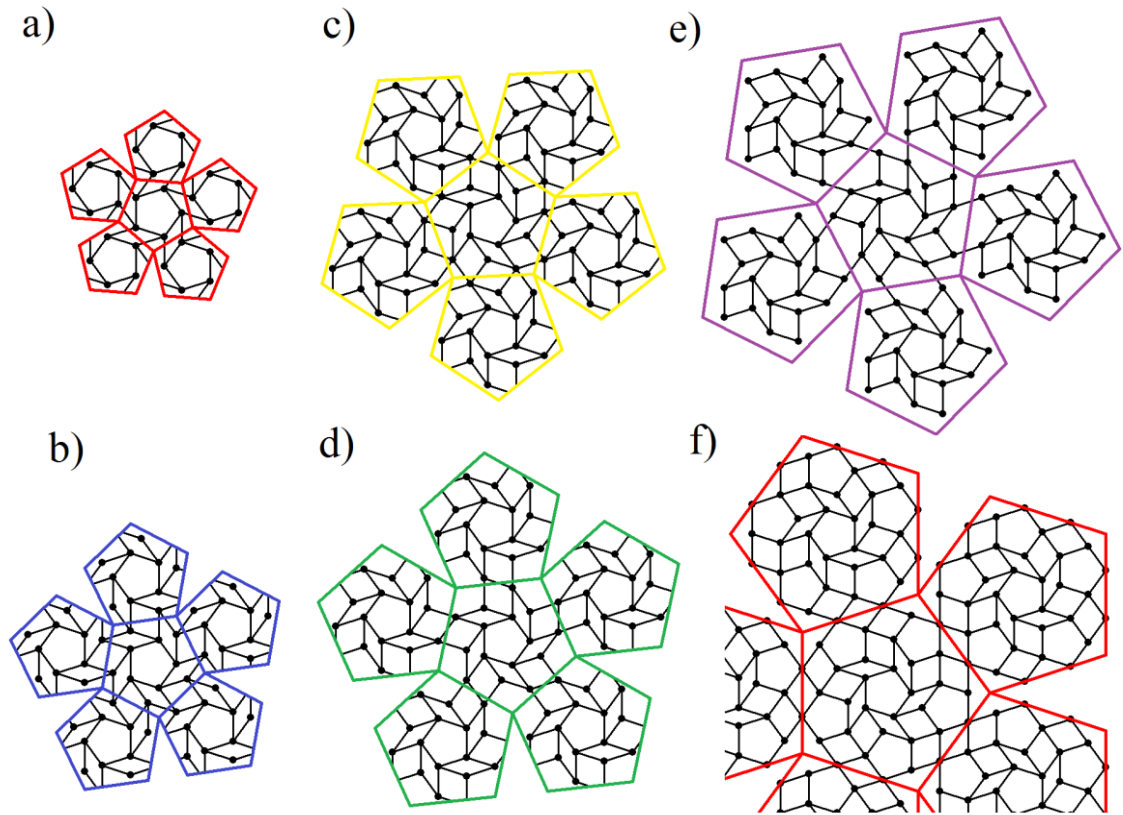

**Рисунок 3** - Развертки шести капсидов, соответствующих рисунку 2(b).

Заметим, что при склейке развертки в додекаэдр и его выпучивании в соответствие с принципом квазиэквивалентности КК должна происходить некоторая дополнительная симметризация структуры. Иначе расстояния, равные на плоскости, станут сильно неравными на сфере и из плоской додекагональной развертки не получить сферический тайлинг, состоящий из минимального набора примерно одинаковых структурных единиц и удовлетворяющий таким образом принципу квазиэквивалентности КК.

Рисунок 4 показывает данные симметризованные сферические структуры (смотри крайнюю правую колонку), обладающие икосаэдрической симметрией *I* и соответствующие хорошо известным вирусным капсидам (смотри крайне левую колонку). Первый, четвертый и пятый приведенный капсид удовлетворяют геометрической теории Каспара - Клуга, а второй, третий и шестой – нет, так как соответствующие структуры не представимы в виде пентамеров и гексамеров. В центральной колонке находятся рисунки, демонстрирующие примерное расположение определенных нами центров тяжести белковых молекул в данных капсидах. Для капсида БПВ приближенные координаты центров тяжести белковых молекул были определены нами ранее на основании функции плотности их распределения [6]. Для остальных показанных капсидов положения центров тяжести протеинов взяты из [8,9].

Сборка показанных на рисунке 3 разверток, в додекаэдр с последующим его выпучиванием и симметризацией узлов возникающих трехмерных объектов, очевидно, проводит к структурам капсидов, показанным на рисунке 4. При этом в некоторых случаях смещения позиций при симметризации оказываются весьма существенными. Например, сильно искаженный гексамер, лежащий на глобальной икосаэдрической оси второго порядка при переходе от рис. 3(e) к рис. 4(e) становится практически правильным, а разница между тонкими и толстыми ромбами при склейке и выпучивании развертки (сравни рис. 3(d) и рис. 4(d)) практически исчезает.

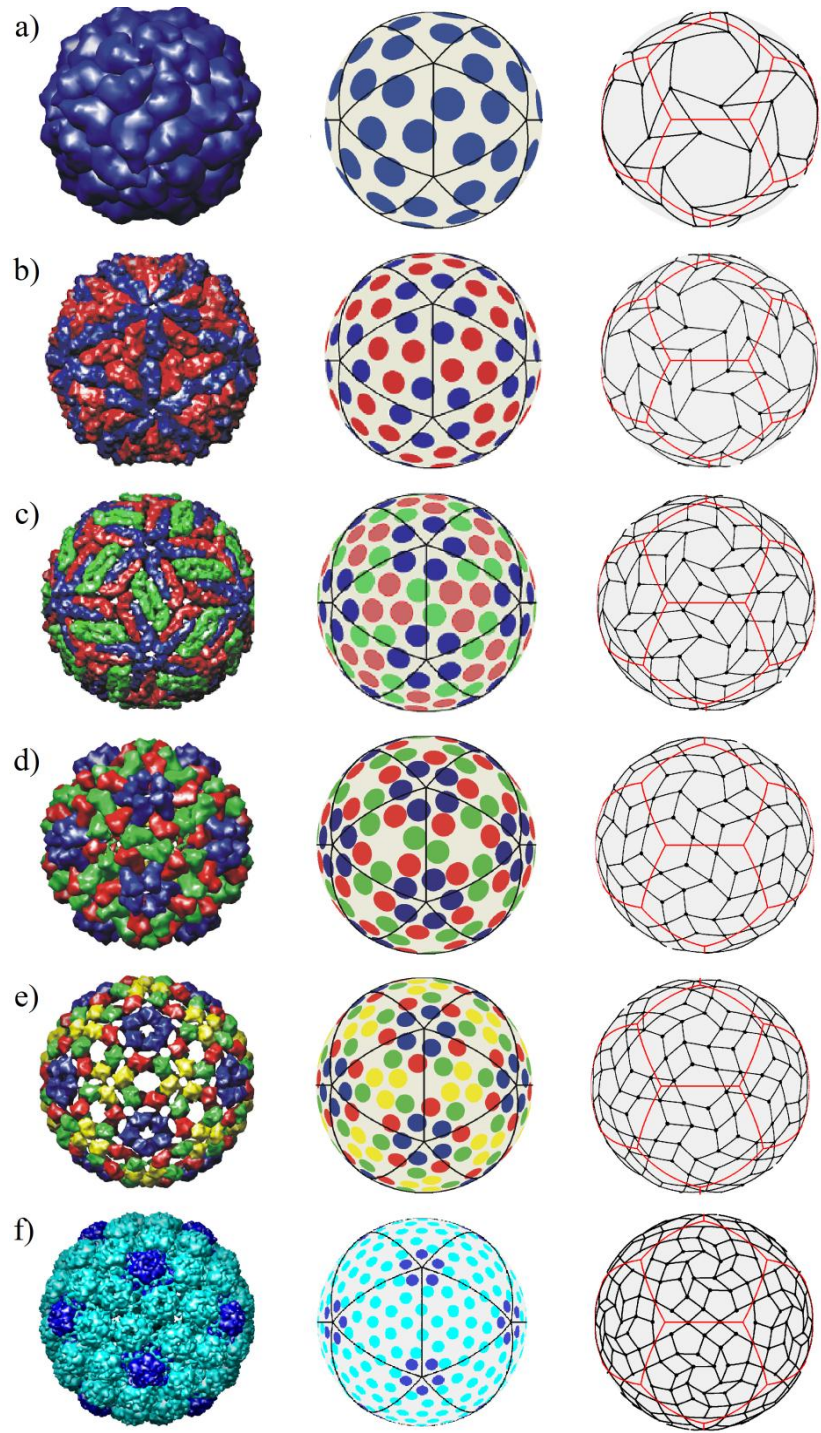

**Рис 4.** Структуры капсидов малых вирусов и регулярные сферические тайлинги с икосаэдрической симметрией *I*: (a) - сателитный вирус табачной мозаики *T*=1; (b) - L-A вирус *T*=2; (c) - вирус лихорадки *T*=3; (d) - вирус хлорозомы вигны *T*=3; (e) - Синдбис вирус *T*=4; (f) - вирус бычьей папилломы *T*=6.

Другой интересный пример применимости развиваемой теории – сдвоенный капсид вируса кукурузного стрика из семейства Близнецов (смотри рис. 5(a), адоптированный из [10]). Сдвоенный дефектный (*T*=1) капсид данного вируса состоит из 22 пентамеров организованных из 110 белков и может быть получен в рамках нашей теории путем построения развертки сдвоенного додекаэдра. При этом каждая грань развертки абсолютно идентична грани простейшей додекаэдрической развертки, показанной на рис. 3(a). Подчеркнем, что если исходить из икосаэдрической геометрии капсида, то вообще не понятно каким образом нужно проводить соединение двух икосаэдрических капсидов с *T*=1. Додекаэдрическая геометрия дает решение этой проблемы. Соединяя два додекаэдра таким образом, чтобы они имели общую грань, мы приходим к развертке (смотри рис. 5(a)) структуры данного капсида, показанной на рис. 5(b). После склейки и выпучивания обеих дефектных T=1 капсидов, получается трехмерная модель структуры капсида, показанная на панели (c).

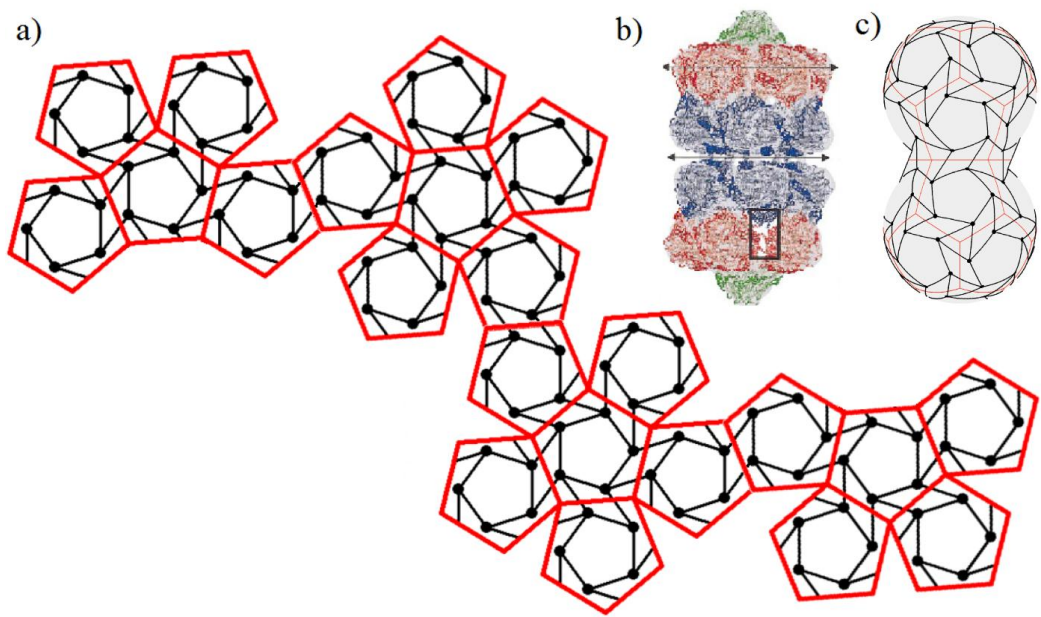

Рис. 5. Сдвоенный капсид вируса кукурузного стрика из семейства Близнецов. a) Развертка данного капсида, основанная на додекаэдрической геометрии. b) Экспериментальная структура, состоящая из 110 капсомеров [10]. c) Идеализированная трехмерная модель структуры капсида вируса, представленная в виде двух сопряженных сферических тайлингов.

## 4. Обсуждение и заключение

В работе построена объединенная теория структур малых (*T*<6) и средних (*T*=6) вирусов, основанная на теории кристаллизации Ландау и подходе волн плотности. В рамках теории Ландау рассматривается образование плоских хиральных структур, заведомо обладающих осью пятого порядка, что позволяет правильно описывать хиральную организацию протеинов в малых и средних сферических капсидах вблизи глобальной оси симметрии 5-го порядка. Получаемый в рамках теории кристаллизации плоский порядок переносится на сферу через развертку додекаэдра. Разработанная в рамках классической квазикристаллической теории концепция нелинейной фазонной деформации, позволяет не только сопрячь плоские структуры, принадлежащие смежным граням додекаэдра, но и объясняет общие закономерности организации протеинов в малых и средних вирусных капсидах.

В отличие от структур малых вирусных капсидов, хорошо описываемых в рамках теории сферической кристаллизации [8,9], структура капсидов семейства Паповавирусов (T=6) не может быть получена в рамках предшествующих работ [8,9]. Предлагаемый в настоящей работе подход одинаково хорошо пригоден как для малых, так и для средних вирусных капсидов. Развитая объединенная теории показывает, что структурный тип БПВ – не экзотическая аномалия, а закономерный последний член ряда капсидных структур с меньшим количеством протеинов. Именно изучение малых и средних вирусных капсидов с позиций додекаэдрической геометрии позволяет найти общие закономерности устройства этих структур. В отличие от икосаэдрической развертки, додекаэдрическая развертка приводит к образованию 20 (а не 12) формальных топологических дефектов, возникающих в процессе переноса плоского порядка на сферу. Данные формальные дефекты оказываются расположенными на осях третьего (а не пятого) порядка глобальной икосаэдрической симметрии. Возможность использования додекаэдрической развертки для исследования структур малых вирусов означает, что пентамеры в практически сферической структуре соответствующих капсидов нельзя рассматривать в качестве дефектов. Таковыми их делает лишь применение икосаэдрической развертки к малым капсидам. Для таких капсидов пентамеры являются не дефектами, а закономерными структурными элементами. Наконец развертка додекаэдрического типа оказывается наиболее разумным способом объяснить с феноменологической точки зрения структуры сдвоенных капсидов семейства Близнецов.

Необходимость сопряжения порядка на смежных гранях додекаэдрической развертки делает развиваемую в работе теорию несколько более сложной, чем теория кристаллизации на сферической поверхности [8,9]. Однако теория плоской кристаллизации имеет несколько более широкую область применения, чем подход [8,9]. Область применения данной теории можно было бы еще более расширить, рассмотрев в дальнейшем теорию плоской кристаллизации структур, заведомо обладающих осью третьего порядка. Наряду с малыми капсидами (T<6) такая теория, фактически рассматривающая соразмерные плоские волны на поверхности икосаэдра, должна описывать и большие капсиды (T>6), характеризуемые икосаэдрической геометрией и гексагональным порядком протеинов внутри граней.